\documentclass[10pt,letterpaper]{article}
\usepackage[top=0.85in,left=1.in,footskip=0.75in,marginparwidth=2in]{geometry}

\usepackage[utf8]{inputenc}

\usepackage{cite}

\usepackage{nameref}

\usepackage[right]{lineno}

\usepackage{microtype}
\DisableLigatures[f]{encoding = *, family = * }

\raggedright
\usepackage{changepage}

\usepackage[aboveskip=1pt,labelfont=bf,labelsep=period,singlelinecheck=off]{caption}

\makeatletter
\renewcommand{\@biblabel}[1]{\quad#1.}
\makeatother

\usepackage{lastpage,fancyhdr,graphicx}
\usepackage{epstopdf}
\fancyheadoffset[L]{2.25in}
\fancyfootoffset[L]{2.25in}

\usepackage{color}

\definecolor{Gray}{gray}{.25}

\usepackage{graphicx}

\usepackage{sidecap}

\usepackage{wrapfig}
\usepackage[pscoord]{eso-pic}
\usepackage[fulladjust]{marginnote}
\reversemarginpar

\begin{document}
\vspace*{0.35in}

\begin{flushleft}
{\Large\textbf{
Single-shot temporal profile measurement of a soft X-ray laser pulse}
}
\vspace{.35in}
\\
F. Tissandier\textsuperscript{1,*},
J. Gautier\textsuperscript{1},
A. Tafzi\textsuperscript{1},
J.-P. Goddet\textsuperscript{1},
O. Guilbaud\textsuperscript{2},
E. Oliva\textsuperscript{3},
G. Maynard\textsuperscript{2}
P. Zeitoun\textsuperscript{1},
S. Sebban\textsuperscript{1}
\\
\bigskip
\textbf{1} Laboratoire d'Optique Appliqu\'ee, ENSTA-Paristech, CNRS, Ecole Polytechnique, Universit\'e Paris-Saclay, 828 Bv des Mar\'echaux,91762 Palaiseau, France
\\
\textbf{2} Laboratoire de Physique des Gaz et des Plasmas, CNRS, Universit\'e Paris-Sud bat. 210, 91405 Orsay, France
\\
\textbf{3} Instituto de Fusi\'on Nuclear, Universidad Polit\'ecnica de Madrid, Spain
\\
\bigskip
* fabien.tissandier@ensta-paristech.fr

\end{flushleft}

\section*{Abstract}
We report an original method allowing to recover the temporal profile of any kind of soft X-ray laser pulse in single-shot operation. We irradiated a soft X-ray multilayer mirror with an intense infrared femtosecond laser pulse in a traveling wave geometry and took advantage of the sudden reflectivity drop of the mirror to reconstruct the temporal profile of the soft X-ray pulse. We inferred a pulse shape with a duration of a few ps in good agreement with numerical calculations and experimental work.

\section*{}
Intense and short soft X-ray light pulses offer unprecedented possibilities for studying ultrafast phenomena in matter. On the one hand, X-ray free-electron lasers (FELs) are user-dedicated facilities providing multi-millijoule femtosecond pulses, and on the other hand there is a wide range of laboratory-size soft X-ray sources such as high-order harmonics or plasma-based soft X-ray lasers (SXRLs) covering the whole soft X-ray spectral range with pulse durations from the attosecond to the picosecond.

Measurement of the pulse duration is an essential part of a soft X-ray source characterization and is crucial for a number of applications. Although using X-ray streak cameras is a convenient way to achieve this, their single-shot resolution is usually limited to 1~ps. In order to increase the resolution, several techniques based on the autocorrelation of the pulse, or on the cross-correlation of the pulse with a given ultrafast gating phenomenon, have been demonstrated. The autocorrelation method records the signal coming from two photo-ionization of He or Ar gas\cite{kobayashi96}\cite{kobayashi98}\cite{sekikawa02} when the main soft X-ray pulse and its replica are temporally overlapped. This method is very straightforward and the resolution is limited by the soft X-ray pulse duration. However, this method does not distinguish between the front and the rear of the soft X-ray pulse, which can be of high importance for applications performed using asymmetrical pulses.

Among the cross-correlation methods, is the laser-assisted photoelectric effect\cite{schins94}\cite{bouhal97}\cite{hentschel01}. When the soft X-ray pulse and the laser pulse are temporally overlapped on the target, some changes such as sidebands or peak energy shift appear on the emitted photoelectron spectrum. Recording these changes while scanning the time delay between the two pulses constitutes a soft X-ray pulse duration measurement with a resolution limited by the laser pulse duration. An alternative method is based on rapid change of the inner-shell resonant absorption in the soft X-ray region of a gas undergoing optical field ionization\cite{oguri01}\cite{oguri04}\cite{oguri04b}. The population change from neutral to ionized is faster than the laser pulse rise time, allowing this method to reach resolutions shorter than the laser pulse duration. Finally, the high number of photons per pulse provided by FELs allowed to use the soft X-ray pulse as a pump and induce infrared reflectivity\cite{maltezopoulos08} or transmission\cite{riedel13} change in a selected sample. These experiments were implemented in a single-shot geometry, leading to pulse duration measurements with a resolution shorter than the pump pulse duration. It is also to be noticed that, when able to achieve a measurement in a single shot, most of these methods require previous knowledge of a reference measurement.

The technique we report here is suited to soft X-ray pulses which are not intense enough to induce ultrafast changes of optical properties in matter, and is based on the rapid reflectivity drop of a soft X-ray multilayer mirror following irradiation by an intense, 30~fs infrared pulse. The soft X-ray laser pulses are generated by seeding an optical-field-ionized Ni-like Kr soft X-ray amplifier at a wavelength of 32.8~nm\cite{sebban02} with a high-order harmonic beam of the driving laser\cite{zeitoun04}\cite{depresseux15}. The amplifier is generated by focusing 20~TW infrared pulses into a cell filled with krypton. The harmonic seed is generated in an argon-filled cell using another beam of the same laser system, and refocused into the amplifier. In this configuration, the amplifier electron density is around $6\times10^{18}$~cm$^{-3}$ and the gain duration as well as the SXRL pulse duration are expected to be of a few ps\cite{guilbaud10}.

\begin{figure}[htbp]
\centering
\fbox{\includegraphics[trim=2cm 0 3cm 0,clip=true,width=.9\linewidth]{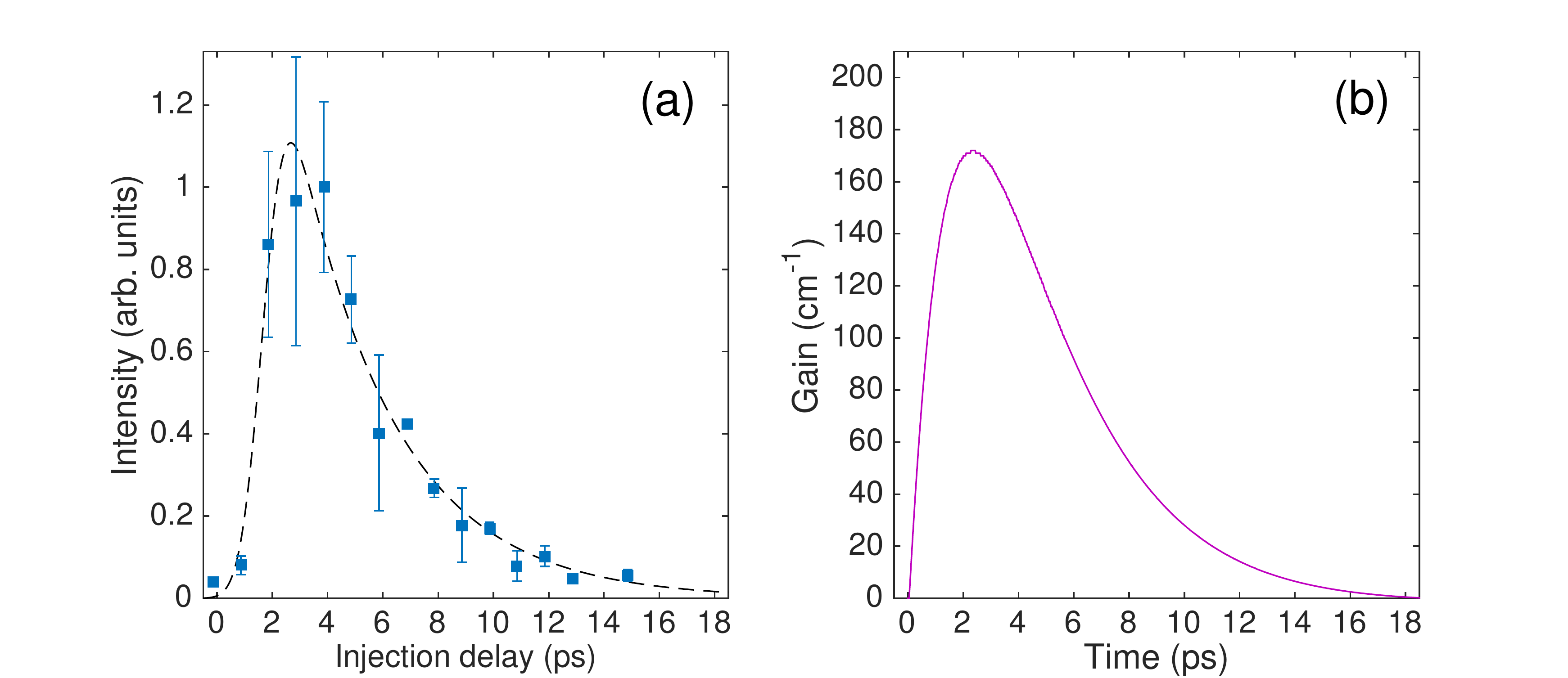}}
\caption{Gain dynamics of the soft X-ray amplifier: (a) the seeded soft X-ray laser intensity is measured as a function of the injection time of the seed in the amplifier. Each data point results from the average over 5 shots. The dotted line comes from an exponentially-modified Gaussian fitting and shows a duration of 5~ps. (b) is the calculated gain using a time-dependent collisional-radiative code.}
\label{gain}
\end{figure}

Figure~\ref{gain}-a) shows the measured soft X-ray intensity as a function of the delay between the creation of the amplifier and the injection of the harmonic seed. This represents a measurement of the gain temporal dynamics at the amplifier's given density. The gain rises rather sharply due to efficient pumping of the population inversion in the amplifier by collisions between hot electrons and Ni-like Kr ions. The main terminating mechanism is quenching of the lasing ions due to collisional ionization and thus occurs on a picosecond time scale related to the characteristic collision time at the amplifier electron density. We can estimate from this data that the soft X-ray gain duration is around 5~ps FWHM. This duration is in good agreement with the data shown on Figure~\ref{gain}-b) which represents a time-dependent calculation of the soft X-ray amplifier gain within the same parameters. We used the collisional-radiative OFI-0d code\cite{cros06} that successively computes the OFI of neutral krypton and the evolution of the Ni-like Kr plasma through collisions.

\begin{figure}[htbp]
\centering
\fbox{\includegraphics[width=\linewidth]{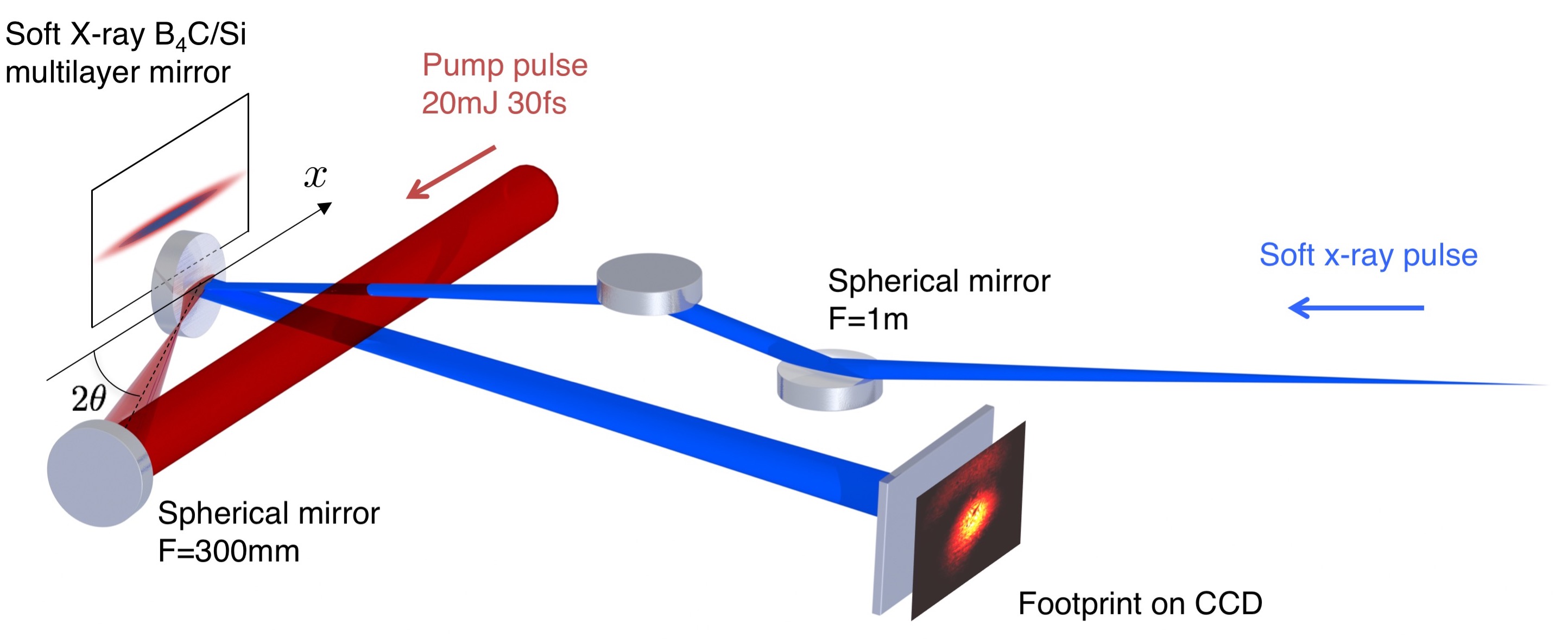}}
\caption{Setup of the experiment. The soft X-ray mirror is pumped in a traveling wave geometry using an oblique incidence spherical mirror. The soft X-ray beam is line-focused on the mirror using a grazing-incidence spherical mirror. A soft X-ray CCD camera records the footprint of the reflected beam at a distance where its symmetrical Gaussian profile is recovered.}
\label{setup}
\end{figure}

The experiment was set up as shown on figure~\ref{setup}. The SXRL pulse beam is focused at near-normal incidence on the soft X-ray mirror using a grazing-incidence spherical mirror. The multilayer mirror is set at the tangential focal plane such that the soft X-ray focal spot on the mirror is a 3mm-long horizontal line. The mirror is composed of 18 B$_4$C/Si layers ensuring a reflectivity of 25\% at 32.8~nm at the considered angle. A soft X-ray CCD camera is used to record the footprint of the reflected SXRL beam.

The infrared pump beam is composed of 20~mJ, 30~fs pulses from the same Ti:Sa laser system, which ensures the measurement is entirely jitter-free. The pump intensity necessary to ensure a total extinction of the soft X-ray reflectivity was found to be lower than $10^{14}$~W/cm$^2$. The beam is focused using a spherical mirror at an incidence angle  $\theta=15^\circ$, producing a 8mm-long line focus on the multilayer mirror. This technique causes the pump beam wavefront to be tilted with respect to the mirror surface, allowing the energy deposition on the mirror to travel along the line focus. The corresponding spatial coordinate $x$ in the multilayer mirror plane thus corresponds to a temporal delay $\tau=x\cos(2\theta)/c$ between the arrivals of the pump and SXRL pulses. Since the multilayer mirror is reflective before pump energy deposition and absorbent when damaged, the reflected soft X-ray intensity profile on the mirror results from an integration of the temporal profile of the soft X-ray pulse :
\begin{equation}\label{eq1}
I(x=\frac{c\tau}{\cos2\theta})=I_0(x)\times\int_{-\infty}^{\tau}\hat{\phi}(t)dt
\end{equation}
where $\hat{\phi}$ is the normalized temporal intensity profile of the soft X-ray pulse. $I_0(x)$ represents the envelope of the signal we seek to recover and is given by the transverse intensity profile of the SXRL beam without pump beam. With a suitable numerical processing, the temporal profile of the SXRL pulse can thus be retrieved. It is also to be noted that in this setup, the maximum pulse duration that can effectively be probed is given by the time span $\Delta\tau$ related to the spatial extension of the soft X-ray focal spot along the $x$ direction:
\begin{equation}
\Delta\tau=\Delta x \frac{\cos(2\theta)}{c}
\end{equation}
where $\Delta x $ represents the spatial extension of the soft X-ray beam. Here, this time span is 12~ps which should be enough considering the expected pulse duration. Should the pulse be longer, only the first 12~ps can be probed without distorting the measurement.

The temporal resolution of the experiment as well as the accurate synchronization of the pump pulse with the SXRL pulse were determined by substituting the SXRL pulse with the 30~fs infrared pulse used to drive the high-harmonic generation. This probe pulse is accurately synchronized with the SXRL laser pulse. In this case, due to the short duration of both the pump and probe pulses, a sharp knife-edge clipping of the probe beam is expected in place of the above-mentioned modulation. The synchronization of the two pulse can be precisely adjusted by altering the amount of clipped beam while the resolution is given by the width of the clipping function. We measured a resolution of 500~fs, well above the 30~fs pump pulse duration. The degradation of the mirror reflectivity is caused by the OFI of the mirror surface\cite{teubner01} by the pump pulse and thus expected to be faster than 500~fs. This is likely to be caused by diffraction on the damaged mirror and propagation to the CCD.

\begin{figure}[htbp]
\centering
\fbox{\includegraphics[width=.8\linewidth]{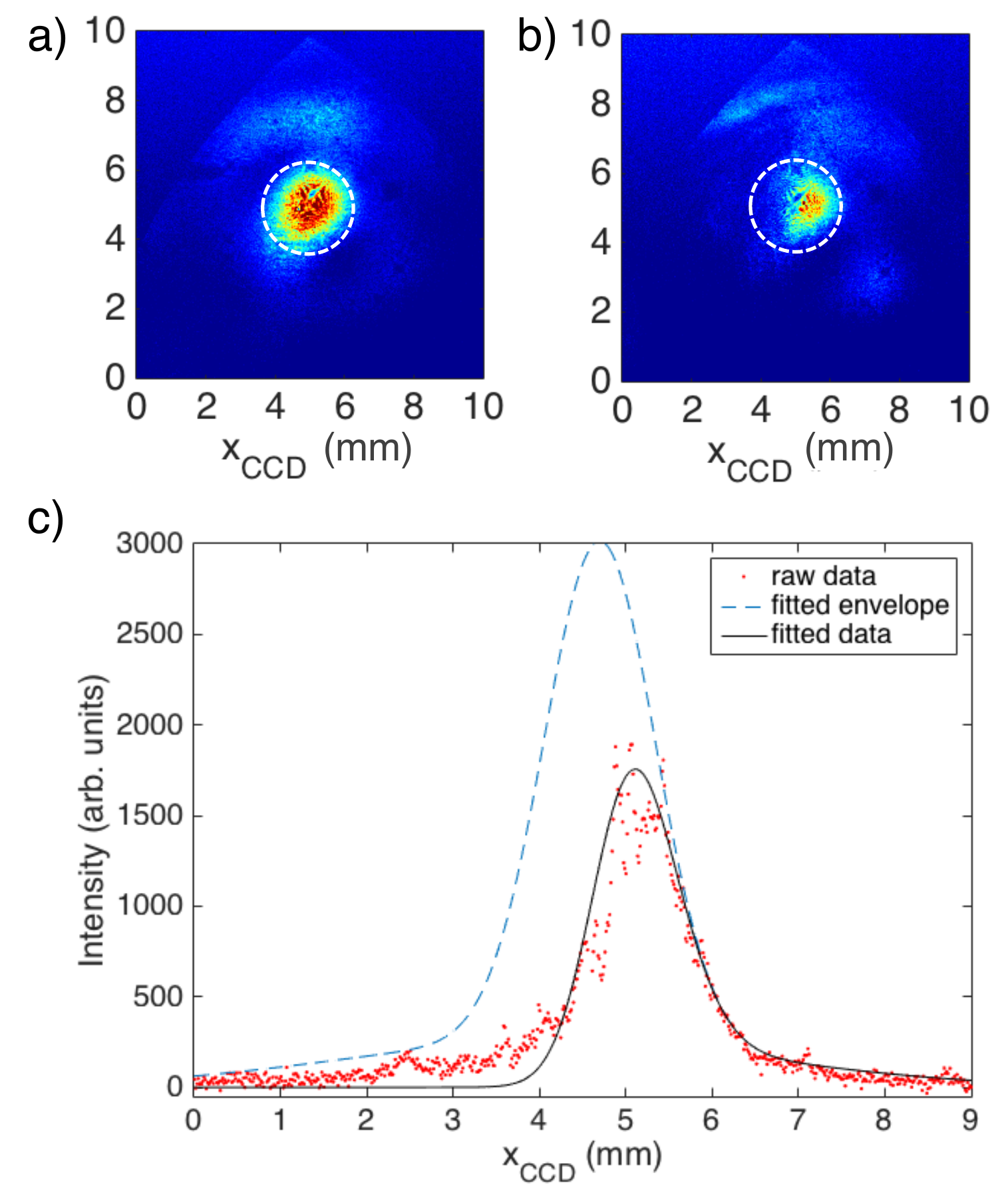}}
\caption{Footprint of the SXRL laser beam reflected on the a) undamaged and b) damaged mirror. The dashed circle is a guide to the eye. The beam horizontal cross-sections are plotted in c). The dashed line represents the fitted intensity profile of the reference beam (intact mirror) while the dots and plain line represent respectively the data points and fitted curve of the intensity profile of the beam reflected on the damaged mirror. The implicit time axis is oriented along positive $x$.}
\label{footprintdata}
\end{figure}

Figure~\ref{footprintdata} shows the recorded footprint of the reflected SXRL beam on (a) the undamaged mirror and (b) the damaged mirror, in presence of the synchronized pump pulse. Since the CCD camera is relatively far from the grazing incidence spherical mirror, the footprint recorded after reflecting on an intact multilayer mirror is circular rather than oval. As expected, the SXRL footprint is smoothly clipped on its left, corresponding to where the pump energy first hits the mirror. Accordingly, the implicit time axis in this figure is oriented along positive $x$. As suggested by Eq.~\ref{eq1}, the SXRL pulse temporal profile $\hat{\phi}(t)$ can be extracted from the cross-section of this spatial profile along the $x$ (horizontal) direction. The corresponding data points were plotted on Figure~\ref{footprintdata}-c) (red dots). The cross section was integrated over 4 pixels in the vertical direction, to ensure smoothing of the resulting curve while not degrading the resolution.

To retrieve the SXRL temporal profile $\hat{\phi}(t)$, this data has to be divided by the envelope $I_0(x)$ and then differentiated with respect to $\tau$. However, due to the noisy nature of $I(x)$ and $I_0(x)$, computing these operations --in particular derivation-- using straightforward techniques amplify the noise so much that the result is practically useless. We chose to fit both sets of experimental data. The reference beam spatial profile $I_0(x)$ is the transverse intensity profile of the SXRL beam without pump. In our experimental setup, the unperturbed footprint of the SXRL beam thus has to be recorded prior to the actual measurement. It was accurately fitted using a sum of two Gaussian functions (dashed line on Fig.~\ref{footprintdata}-c)). To compute the fit of the perturbed SXRL beam data $I(x)$, we assumed an exponentially-modified Gaussian temporal profile $\hat{\phi}(t)$:
\begin{equation}\label{eq2}
\hat{\phi}(t)=\phi_0\exp\left[-\left(\frac{t-t_g}{\sqrt{2}w}\right)^2\right]\mathrm{erfcx}\left[\frac{1}{\sqrt{2}}\left(\frac{w}{T}-\frac{t-t_g}{w}\right)\right]
\end{equation}
where $\displaystyle\mathrm{erfcx}(x)=e^{x^2}\left(1-\frac{2}{\sqrt{2}}\int_x^\infty e^{-u^2}du\right)$ is the scaled complementary error function. The fit function is computed at each data point $x$ on the CCD by integrating $\hat{\phi}$ and multiplying by the envelope function value at the considered point. Fitting was performed using the resulting function on the Gaussian component width $w$ and center $t_g$ and the exponential decay $T$. The fitted curve is plotted on Figure~\ref{footprintdata}-c) (plain line).

The fitted curve diverges from experimental data on the left side (smaller $x$) of the intensity profile. This is very unlikely due to a smooth rising of the SXRL pulse but rather the consequence of diffusion on the damaged mirror. The reflectivity drop cannot be effectively considered as total.

\begin{figure}[htbp]
\centering
\fbox{\includegraphics[width=.8\linewidth]{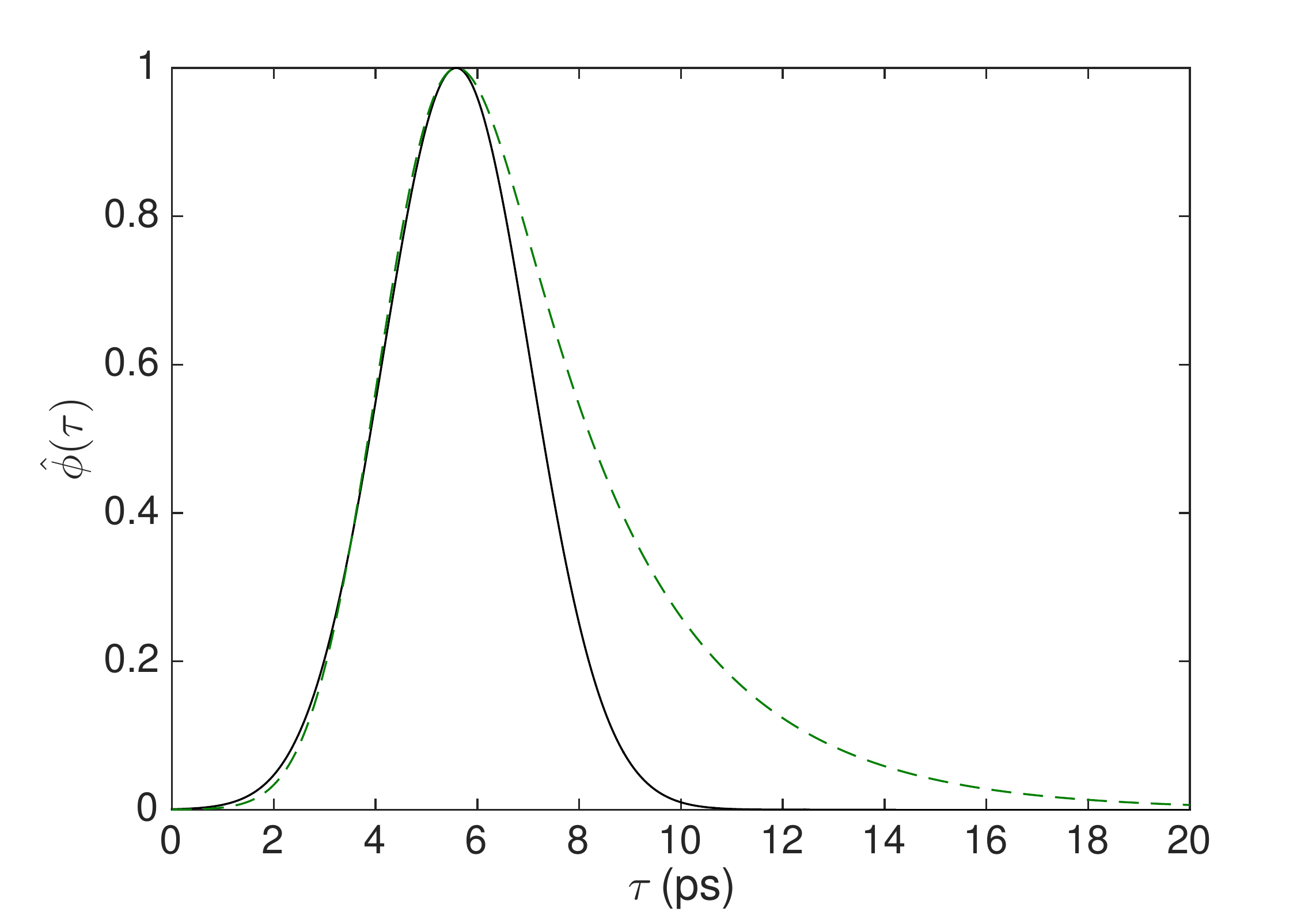}}
\caption{Reconstructed normalized temporal profile of the SXRL pulse. The plain line comes from the best fit presented on Fig.~\ref{footprintdata}. The dashed line results from an asymmetrical fit.}
\label{temprof}
\end{figure}

Once the fit parameters are known, the normalized temporal intensity profile can easily be extracted. It was plotted on Figure~\ref{temprof} (plain line). Since $\tau$ represents the delay between the SXRL and pump pulses, the origin is arbitrary. It exhibits a rising edge of about 3~ps in good agreement with the gain measurements and calculations presented on Fig.~\ref{gain}. Its shape is roughly symmetrical and its FWHM duration is 3.5~ps, a little shorter than previously measured. However, we also plotted (dashed line) on Figure~\ref{temprof} the temporal profile extracted from a different fit forcing a more asymmetric profile, closer to the data shown in Fig.~\ref{gain} . As a result, the rising edge is practically the same, but the following decrease is slower, providing a FWHM pulse duration of 4.5~ps, in better agreement with the data presented in Fig.~\ref{gain}. The uncertainty in the falling edge, and thus in the final pulse duration is due to the fact that this part of the pulse is much harder to ascertain than the rising edge. At a given spatial coordinate $x$ corresponding to a time delay $\tau$ between the pump and SXRL pulses, all the energy contained in the SXRL pulse, from its rise to the instant $\tau$, is reflected on the mirror and sent to the CCD. Therefore, for longer delays, most of the pulse energy is actually collected on the CCD. The intensity measured on the CCD at $x$ is thus expected to be closer to the unperturbed beam intensity $I_0(x)$ (envelope). Since smaller variations from the envelope are harder to discern, this method is intrinsically less accurate at extracting the back part of the pulse. Extraction of the back of the pulse is further complicated by the fact that the relevant delays correspond on the mirror to the intensity drop of the Gaussian spatial profile. This is however not inherent to our method, but rather the consequence of a not-so-ideal match between the transverse (spatial) and longitudinal (temporal) dimensions of the SXRL pulse. Its spatial dimension (around 4~mm) correspond to a time spread of 13~ps, which is a little short to resolve the whole pulse.

In summary, we have proposed and demonstrated an original method aiming at extracting the temporal intensity profile of a soft X-ray pulse in a single-shot using a relatively straightforward experimental setup. The geometry of this setup needs to be adapted to the expected pulse duration, since it will determine the maximum pulse duration that can be effectively probed. We were able to recover the temporal profile of a picosecond duration soft X-ray laser pulse in good agreement with previous work. Since based on the reflectivity drop of a multilayer mirror irradiated by a moderately intense infrared pulse, this method is compatible with any kind of emission in the soft X-ray range and beyond, with the assistance of a moderately intense pump beam ($10^{14}$~W/cm$^2$). The simultaneous measurement of the unperturbed reference beam will be an asset in mitigating shot-to-shot fluctuations of the source. Moreover, imaging the soft X-ray line focus in the mirror plane will alleviate diffraction effects and subsequent blurring on the sensor, significantly increasing the time resolution and opening to the path to ultrafast soft X-ray measurements.

\section*{Acknowledgments}

This work is supported by "Investissements d'Avenir" Labex PALM (ANR-10-LABX-0039-PALM) and has received funding from the European Union's Horizon 2020 research and innovation program under grant agreement No. 654148 Laserlab-Europe.

\end{document}